\documentclass{aa}

\usepackage{graphicx}
\usepackage{txfonts}
\usepackage{natbib}
\usepackage{fnpara} 
\bibpunct[,]{(}{)}{;}{a}{}{,}
\begin{document}

   \title{Are short $\gamma$-ray bursts collimated?\\
          GRB 050709, a flare but no break}
   \titlerunning{A flare but no break in GRB 050709}
   
   \author{D.~Watson, J.~Hjorth,
          P.~Jakobsson, D.~Xu, J.~P.~U.~Fynbo,
          J.~Sollerman, C.~C.~Th\"one and K.~Pedersen
          }
   \authorrunning{Watson~et~al.}

   \offprints{D.~Watson}

   \institute{Dark Cosmology Centre, Niels Bohr Institute, University of
              Copenhagen, Juliane Maries Vej 30, DK-2100, Copenhagen \O,
              Denmark\\
              \email{darach, jens, pallja, dong, jfynbo, jesper, cthoene, kp @astro.ku.dk}
             }

   \date{Received ; accepted }

  \abstract{From the small sample of afterglow lightcurves of short duration
            $\gamma$-ray bursts (GRBs), the decays are rapid, roughly
            following a power-law in time. It has been assumed that the
            afterglow emission in short GRBs is collimated in jets in the
            same way as in long GRBs. An achromatic break in a short GRB
            afterglow lightcurve would therefore be strong evidence in
            favour of collimation in short GRBs. We examine the optical
            lightcurve of the afterglow of the short GRB\,050709, the only
            short GRB where a jet break has been claimed from optical
            data. We show that (1) the decay follows a single power-law
            from 1.4 to 19 days after the burst and has a decay index
            $\alpha = 1.73_{-0.04}^{+0.11}$, (2) that an optical flare at
            $\sim10$ days is required by the data, roughly contemporaneous
            with a flare in the X-ray data, and (3) that there is no
            evidence for a break in the lightcurve. This means that so far
            there is no direct evidence for collimation in the outflows of
            short GRBs. The available limits on the collimation angles in
            short GRBs now strongly suggest much wider opening angles than
            found in long GRBs.
   \keywords{ gamma-rays: bursts---stars: neutron}}

   \maketitle

\section{Introduction} Great progress has been made in the past year on the
origins of short-duration
\citep[$<2$\,s,][]{1984Natur.308..434N,1992AIPC..265..304D,1993ApJ...413L.101K}
$\gamma$-ray bursts (SGRBs), mostly due to the detection of the first
afterglows of SGRBs at X-ray \citep{2005Natur.437..851G}, optical
\citep{2005Natur.437..859H} and radio \citep{2005Natur.438..988B}
wavelengths. Their detection in galaxies with little star-formation
\citep{2005Natur.437..845F,2005Natur.438..988B,2006ApJ...642..989P,2006A&A...450...87G},
and lack of an associated supernova
\citep{2005ApJ...630L.117H,2005A&A...439L..15C}, is in direct contrast to
long-duration GRBs (LGRBs) which are associated with the deaths of massive stars
\citep{1998Natur.395..670G,2003ApJ...591L..17S,2003Natur.423..847H,2004ApJ...609L...5M}.
In fact, the recent SN-LGRB, SN2006aj/GRB060218 has resulted in a large body
of new data on these objects
\citep{2006astro.ph..3530P,2006astro.ph..3279C,2006astro.ph..3495S,2006astro.ph..3377M,2006astro.ph..3832C,2006ApJ...643L..99M}.
In the past few months considerable data has been garnered on the afterglow
properties of SGRBs. With these recent results, the range of distances to,
and isotropic equivalent energies of, SGRBs has expanded
\citep{2006astro.ph..3282L,2006astro.ph..1455S}.

It is generally assumed that SGRB afterglows have properties similar to the
afterglows of LGRBs
\citep[e.g.][]{2001ApJ...561L.171P,2001A&A...379L..39L,2005Natur.437..845F,2006MNRAS.367L..42P,2006A&A...447L...5C,2005Natur.438..988B}.
The afterglows of SGRBs do show fast, approximately power-law decays in
X-ray and optical wavelengths, but with many strong deviations from a simple
power-law model. These deviations are interpreted as energy injection or
short-term flaring
\citep{2006astro.ph..3282L,2006astro.ph..2541L,2006astro.ph..1455S}. Much
has been inferred about the collimation properties of SGRBs from the
variations from a power-law in a single band
\citep{2005Natur.437..845F,2005Natur.438..988B,2006astro.ph..1455S}. Given the strong flaring
activity now known to exist in SGRB decays, it is reasonable to be cautious
about such inferences. Indeed, in only one case to date has a positive claim
been made for a jet break in an SGRB optical lightcurve, GRB\,050709
\citep{2005Natur.437..845F}.

In this paper we analyse the available data on the spectral and temporal
properties of the afterglow of GRB\,050709, the first SGRB where an optical
afterglow was detected and where a claim for a jet break has been made. We
then examine the limits on jet breaks in other SGRBs and compare the
opening angles of SGRBs with LGRBs.

\section{The optical lightcurve of GRB\,050709}

\begin{figure}
  \includegraphics[angle=-90,width=\columnwidth,clip=]{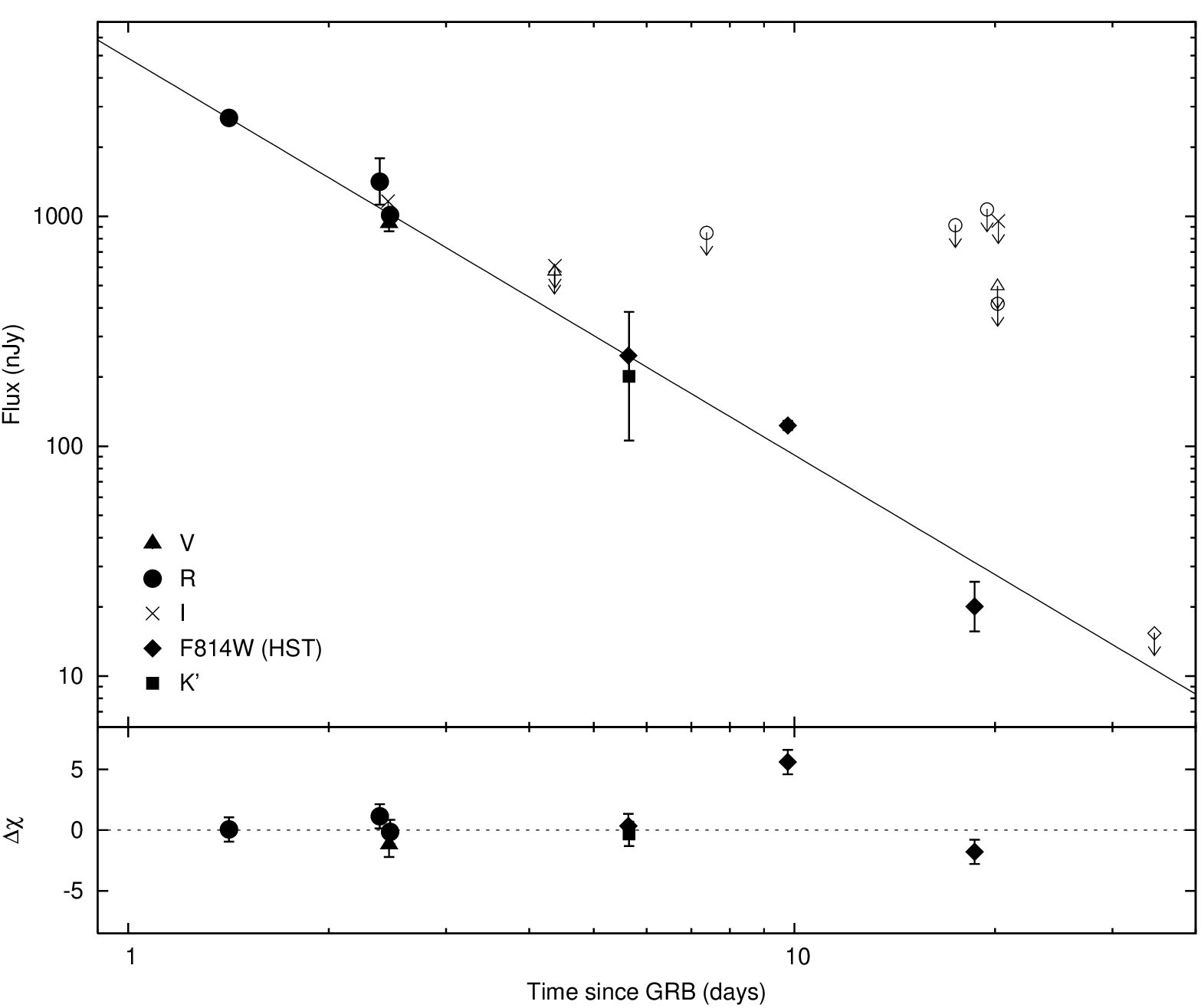}
 \caption{The optical lightcurve of the short GRB\,050709. A single
           power-law decay ($\alpha_{\rm O} = 1.73$) has been fit to the data
           excluding the second HST detection at 9.8 days. An acceptable fit
           is obtained only when this datapoint is excluded.  Data from
           different bands have been corrected to the R-band flux using the
           best-fit power-law spectrum. The fit subtracted from the data
           (residuals) in units of $\Delta\chi$ are plotted in the lower
           panel.}
           \label{fig:opticallc} 
\end{figure}

The optical--near-infrared lightcurve of the afterglow in any one band is
sparsely sampled. But detections have been made in the $V$, $R$, F814W, and
$K^\prime$ bands (Table~\ref{tab:opticaldata}), so we can create a reasonably
sampled lightcurve over a long timescale with a little knowledge of the
broadband spectrum. The $R$ and F814W bands are the best-constrained data
and drive a power-law fit to the data (Fig.~\ref{fig:opticallc}).
Fortunately, these bands are spectrally close, so that the colour-correction
is small.

In long duration GRBs the afterglow continua are predominantly power-laws
\citep[e.g.][]{1998ApJ...497L..17S,2004A&A...427..785J,2004MNRAS.349...31W}.
It seems reasonable that the optical/NIR spectrum of GRB\,050709 can be
represented by a power-law shape especially over the small spectral range
that dominates the lightcurve fit ($R$ to F814W). Using the
near-simultaneous $V$ and $R$ (2.4 days), and F814W and $K^\prime$ (5.6
days) observations, the spectral index of the power-law ($F_\nu \propto
\nu^{-\beta}$) was $\beta_{\rm O} = 1.7\pm 0.8$, and $\beta_{\rm O} = 1.2\pm0.7$,
respectively. Combining these data gives $\beta_{\rm O} = 1.4\pm0.5$. The upper
limit in the $I$ band at 2.4 days is consistent with this spectral index.
This is bluer than the $\beta_{\rm O} = 2.3\pm0.7$ derived by
\citet{2006A&A...447L...5C}, but still within the $1\sigma$ error bounds.
All detections before 5 days in the literature have assumed a zero flux from
the afterglow at about a week. To correct for this, a small flux derived
from the late afterglow (using the HST lightcurve) was added to the early
flux values. The offset added to the early data is partly responsible for
the bluer spectral index derived here.

Using this power-law spectrum with $\beta_{\rm O}=1.4$, the data were converted to
fluxes at the effective wavelength of the $R$ band. The precise value of the
colour correction does not substantially affect the lightcurve; values of
$\beta_{\rm O}$ between 1.0 and 2.4 give very similar results. This relative
insensitivity to the colour correction is because $\beta_{\rm O}$ is derived from
the same wavelength range as the lightcurve data and, as mentioned above,
because the lightcurve fit is driven primarily by the $R$-band and F814W
data, where the wavelength separation is quite small.

\begin{table}
 \caption{Optical observations of GRB\,050709 in 2005. Colour-corrected fluxes
          used in Fig.~\protect\ref{fig:opticallc} are given in column~5.}
 \label{tab:opticaldata}
 \setlength{\tabcolsep}{6pt}
\begin{minipage}{\columnwidth}
  \begin{tabular}{@{}l@{}crccc@{}}
\hline\hline
  \multicolumn{2}{@{}c}{Observation} & \multicolumn{1}{c}{$\Delta t$} & Magnitude & Band & $R$-band Flux\\
  Date & Time & (days)&   &  & \multicolumn{1}{c}{($\mu$Jy)} \\
\hline
  \multicolumn{2}{@{}c}{July}\\
  \footnote{Danish 1.54\,m \citep{2005Natur.437..859H}}11 	& 08:37 & 1.4166 	& $22.71\pm0.06$ 	& $R$	& $2.7\pm0.1$\\
						   $^a$12	& 07:53	& 2.3862	& $23.46\pm0.28$	& $R$	& $1.4^{+0.4}_{-0.3}$		\\
						   $^b$12 	& 09:32 & 2.4551 	& $>23.25$ 		& $I$	& $<1.2$ 					\\
  	     \footnote{VLT \citep{2006A&A...447L...5C}}12	& 09:44	& 2.4635 	& $24.38\pm0.10$ 	& $V$	& $0.93^{+0.08}_{-0.07}$\\
						   $^b$12 	& 09:57 & 2.4725 	& $23.83\pm0.07$	& $R$	& $1.01\pm0.06$ 			\\ 
						   $^b$14 	& 07:21 & 4.3642 	& $>25.00$		& $V$	& $<0.6$ 					\\
						   $^b$14 	& 07:21 & 4.3718 	& $>24.10$ 		& $I$	& $<0.6$ 					\\ 
						   $^c$15 	& 13:49 & 5.6336 	& $25.08\pm0.02$ 	& F814W	& $0.248\pm0.005$ 			\\ 
  \footnote{HST and Subaru \citep{2005Natur.437..845F}}15	& 14:06	& 5.6454 	& $22.1\pm0.7$ 		& $K^\prime$	& $0.2^{+0.2}_{-0.1}$\\ 
						   $^a$17 	& 07:46 & 7.3812 	& $>24.1$ 		& $R$	& $<0.8$ 					\\
						   $^c$19 	& 17:11 & 9.7739 	& $25.84\pm0.05$ 	& F814W	& $0.123\pm0.006$ 			\\ 
						   $^a$27 	& 09:07 & 17.4378 	& $>24.0$ 		& $R$	& $<0.9$ 					\\
						   $^c$28 	& 13:48 & 18.6329 	& $27.81\pm0.27$ 	& F814W	& $0.020^{+0.006}_{-0.004}$ \\[5pt]
						   $^a$29 	& 09:30 & 19.4536 	& $>23.8$		& $R$	& $<1.1$ 					\\
						   $^b$30 	& 02:37 & 20.1669 	& $>25.20$ 		& $V$	& $<0.5$ 					\\
						   $^b$30 	& 02:54 & 20.1787 	& $>25.00$ 		& $R$	& $<0.4$ 					\\ 
						   $^b$30 	& 04:10 & 20.2315 	& $>23.50$ 		& $I$	& $<1.0$ 					\\
  \multicolumn{2}{@{}c}{August}\\
						   $^c$13 	& 15:17 & 34.6947 	& $>28.1$ 		& F814W	& $<0.015$ 					\\ 
\hline 
\end{tabular} 
\end{minipage}
\end{table}

The resulting lightcurve was then fit with a single power-law
(Fig.~\ref{fig:opticallc}), yielding a poor fit regardless of the colour
correction ($\chi^2 = 27.6$ for 6 degrees of freedom). The poor fit was
entirely due to the second HST datapoint at 9.8 days.  A broken power-law
improved the fit slightly, but still the fit was unacceptable ($\chi^2 =
16.2$ for 4 degrees of freedom) and in fact required a flattening rather
than a steepening of the decay. However, excluding the second HST detection
from the dataset allowed a good fit to be obtained with a single power-law
($F(t) \propto t^{-\alpha}$) with a moderately steep decay index
$\alpha_{\rm O}=1.73\pm 0.04$ ($\chi^2 = 6.2$ for 5 degrees of freedom).
Adding the uncertainty from the colour correction gives $\alpha_{\rm
O}=1.73_{-0.04}^{+0.11}$.

\subsection{Comparison with X-rays} The first \emph{Chandra} observation
shows a clear detection of the source with $\sim50$ counts \citep{2005Natur.437..845F}.
Assuming a power-law model with Galactic absorption, the power-law spectral
index is $\beta_{\rm X}=1.6\pm0.3$, consistent with $\beta_{\rm O}$ derived
above. The 0.3--8.0\,keV flux is
$7\pm2\times10^{-15}$\,erg\,cm$^{-2}$\,s$^{-1}$
($2\pm1\times10^{-16}$\,erg\,cm$^{-2}$\,s$^{-1}$\,keV$^{-1}$ or
$0.08_{-0.04}^{+0.06}$\,nJy at 5\,keV). The optical-to-X-ray spectral index
is then $\beta_{\rm OX}=1.2\pm0.1$ at 2.45 days. This spectral index is
consistent with all of the spectral indices derived in the optical/NIR. The
data are therefore consistent with a power-law spectrum with a single
power-law index from the NIR to the X-ray regime.

\section{Implications for short GRBs}
It has already been noted that the HST data are not consistent with a single
power-law decay in GRB\,050709 and it was suggested that the second
detection with HST represented a break to a steeper decay rate, consistent
with a jet break \citep{2005Natur.437..845F}. It is clear from this analysis
of all the available data, that there is no evidence for a break in the
lightcurve of GRB\,050709. The HST datapoint at 9.8 days, instead,
represents a flare or a rebrightening in the optical. This is not surprising
empirically, in light of the probable flare in the X-ray data for this burst
\citep{2005Natur.437..845F} at 16 days, as well as the rebrightenings observed
in other SGRBs: GRB\,050724
\citep{2005Natur.438..994B,2005Natur.438..988B}, GRB\,051210
\citep{2006astro.ph..2541L}, GRB\, 051221A \citep{2006astro.ph..1455S},
GRB\,060121 \cite{2006astro.ph..3282L}.
The X-ray flare may be directly related to the optical
rebrightening, though it would require a slow rise and a very rapid fall if
they were correlated.

The rebrightening observed in the optical/NIR is about two orders of
magnitude below the faint Type\,Ic supernova (SN), 1994I and cannot be fit
with standard SN templates because it requires a much earlier rise-time and
a quicker decay than observed in SNe. If the flares in the optical and X-ray
are associated, this probably also excludes a SN origin for the
rebrightening, since the X-ray flare is so late.

The very late time of these
flares seems to exclude models related to the natural timescale of a
compact-body merger
\citep{2003MNRAS.345.1077R,2004MNRAS.352..753S,2006MNRAS.368.1489O}, as well as those involving
shock heating of a stellar companion unless the orbital distance is much
larger than suggested by \citet{2005astro.ph.10192M}.  Models where the
characteristic spectrum is thermal cannot explain both the X-ray and optical
flares together. The late time of the flaring may also be problematic for
models involving large non-uniformity in the accretion
\citep{2006ApJ...636L..29P,2005ApJ...630L.113K}, since the accretion must
continue for $>10$ days after the burst.

\subsection{Jet breaks in SGRBs lightcurves}

The steep decay reported here could be indicative of a jet-break prior to
the start of optical observations in GRB\,050709, however, the X-ray data
are well-fit (reduced $\chi^2=0.7$) by a single power-law decay from the
HETE-WXM detection of the long soft emission 100\,s after the short burst
\citep{2005Natur.437..855V}, to the late Chandra detection at 16.1 days
(excluding the flare at 16.0 days), with a decay index ($\alpha_{\rm X} =
1.97\pm0.02$) which is close to the optical decay. This indicates that a
break at early times ($<2$ days) is unlikely. In this case, we can limit any
achromatic break to $>10$ days. Indeed, it seems likely that there was no
break as late as the third HST detection at 18.6 days, since the detection
at this time, and in the X-ray at 16 days, are consistent with the single
early power-laws. However we cannot absolutely exclude that such a break
occurred around the time of the flaring, with the flare disguising such a
break. Therefore the conservative limit on any break is $>10$ days. This
limit corresponds to a half opening angle, $\theta_{\rm jet}>23\degr$, using
the relation of \citet{1999ApJ...519L..17S}, an isotropic equivalent energy
$E_{\rm iso}=7\times10^{49}$\,erg \citep{2005Natur.437..845F} and assuming a
density $n=10^{-2}$\,cm$^{-3}$. The limit is not very sensitive to the
assumed density or the derived total energy since the angle is proportional
to $(n/E_{\rm iso})^{0.125}$. The location of GRB\,050709 in a
star-forming galaxy suggests that the density is unlikely to be signicantly
lower than assumed above, a higher density would result in a (slightly)
larger limit on the opening angle. This limit, $\theta_{\rm jet}>23\degr$,
is much larger than the typical opening angle found for LGRBs
\citep{2006ApJ...637..889Z}.

While the lightcurves of SGRBs do decay rapidly, roughly as a power-law,
they are all affected by strong variations, ranging from a moderate
amplitude `wiggling' to very large amplitude flaring \citep[e.g.
GRB\,050709, as noted above, or GRB\,050724,][]{2006astro.ph..3773G}. For
this reason it is difficult to ascertain the decay slope of any underlying
power-law and then fix an achromatic breaktime. This is evidenced by the
first inaccurate suggestions of jet breaks in GRB\,050709 and GRB\,050724
\citep[see Fig.~7 in][]{2006astro.ph..1455S} -- \citet{2006astro.ph..3773G}
report no lightcurve break detected in GRB\,050724 either, out to at least
three weeks after the burst. The break in the X-ray lightcurve of
the afterglow of GRB\,051221A at $\sim5$ days may be a jet break
\citep{2006astro.ph..4320B}. This seems to be consistent with
the available data \citep{2006astro.ph..1455S}, however without strong
limits or detections at other wavelengths to indicate a simultaneous break,
the claim that it is a jet break must be considered weak. The opening angle
of $\sim7\degr$ \citep{2006astro.ph..4320B,2006astro.ph..1455S}
corresponding to a jet break at 5 days must therefore be considered a lower
limit. Evidence of an achromatic break in the lightcurve, critical
to the analysis of the collimation of the outflows of SGRBs, has therefore
yet to be observed in any SGRB.

\begin{figure}
\includegraphics[angle=-90,bb=62 56 558 557,width=\columnwidth,clip=]{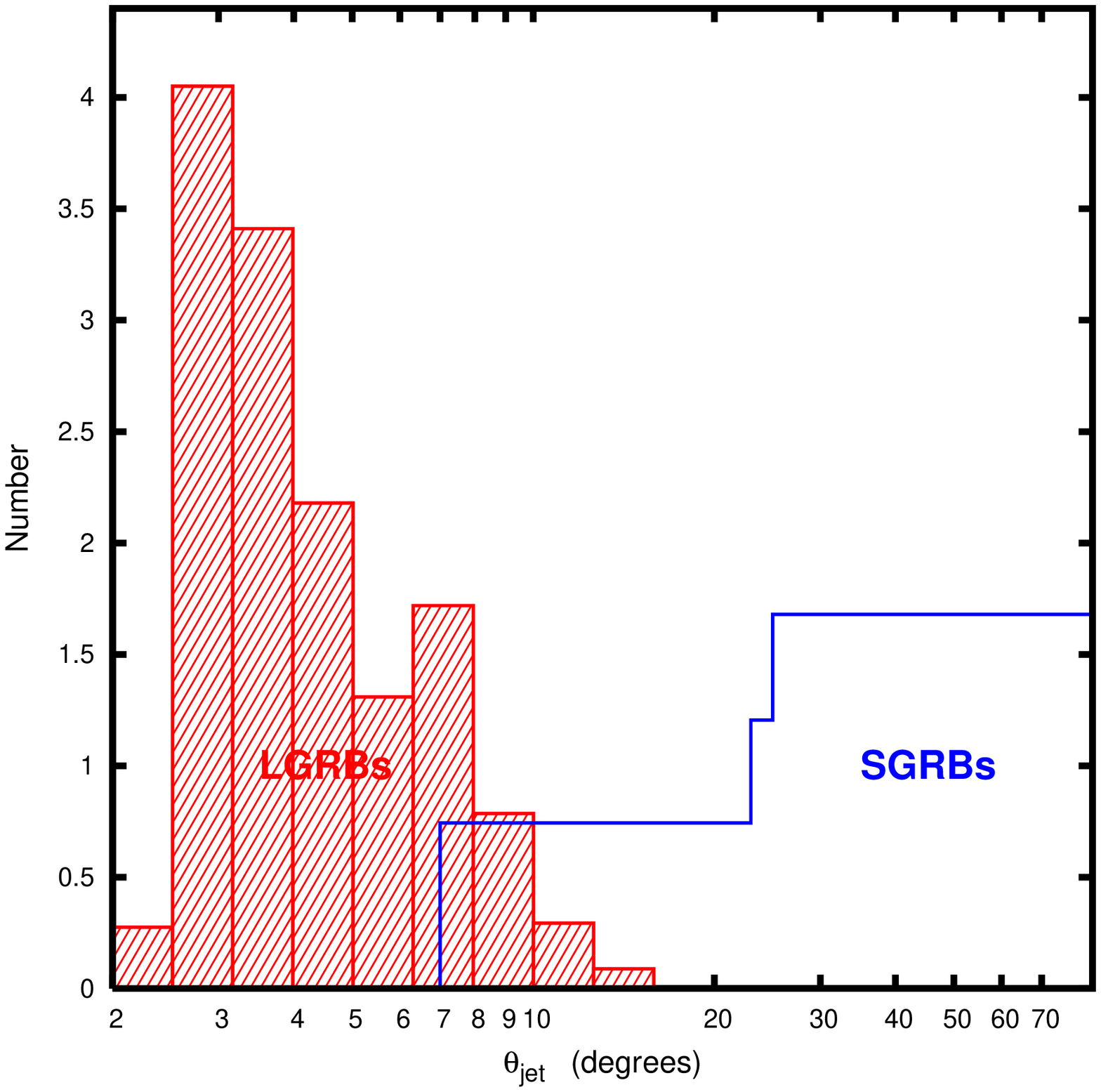}
\caption{Distributions of opening angles for short and long GRBs. Only lower
         limits are available for the short GRBs, but even with only four
         bursts, they seem clearly incompatible with the long GRB
         distribution. The long burst distribution is from the sample of
         \protect\citet{2006ApJ...637..889Z}. The short bursts are
         GRB\,050709 ($>23\degr$, this paper), GRB\,050724
         \protect\citep[$>25\degr$,][]{2006astro.ph..3773G}, GRB\,051221A
         \protect\citep[$>7\degr$,][]{2006astro.ph..1455S} and GRB\,060121
         \protect\citep[$>7\degr$ assuming $z=3$,][]{2006astro.ph..3282L}.
         The distributions have been produced by dividing the probability
         density for each opening angle between the bins using the available
         uncertainties and limits.}
\label{fig:openingangles}
\end{figure}

Lower limits to achromatic break times are now available for two SGRBs with
redshifts (GRB\,050724 and GRB\,051221A) and one without
(GRB\,060121).\footnote{The limits obtained for GRB\,051210, GRB\,050813 and
GRB\,050509B are very weak: none have an optical afterglow detection and
therefore their redshifts are somewhat uncertain, and in all cases, X-ray
emission is well-detected only in the first few hundred seconds after the
burst.} Combining these limits with the limit for GRB\,050709, we can
compare their opening angles with the distribution of opening angles found
for LGRBs (Fig.~\ref{fig:openingangles}). It is immediately apparent that
the distributions are different, with SGRBs having much larger opening
angles, consistent with no collimation at all. While the opening
angles are fairly insensitive to the assumed density, it might be possible
to decrease the lower limits on the opening angle by as much as a factor of
two if the assumed density could be lowered by a factor of about 300.
However, such low densities could be problematic in trying to reproduce the
properties of the afterglows \citep[e.g.][]{2006astro.ph..1455S}, and in the
cases where the GRBs are found within a galaxy, such low densities can
essentially be excluded.

The minimum $\gamma$-ray energies of the three SGRBs with known redshift is
$1.6\times10^{49}$\,erg, $2.8\times10^{49}$\,erg and $1\times10^{49}$\,erg
for GRB\,050709, GRB\,050724 and GRB\,051221A respectively. Their respective
isotropic equivalent energies are $2\times10^{50}$\,erg,
$3\times10^{50}$\,erg and $2.4\times10^{51}$\,erg
\citep{2005Natur.437..845F,2005Natur.438..994B,2006astro.ph..1455S}. These
limits are clearly different from the values found for classical LGRBs
\citep[$10^{50}-10^{52}$\,erg,][]{2006ApJ...637..889Z}. However, the SGRBs
are substantially closer than most of the LGRBs in this sample. A comparison
with the $\gamma$-ray energies of low-redshift LGRBs -- e.g.\ GRB\,980425
\citep{1998Natur.395..670G}, 020903 \citep{2004ApJ...602..875S}, 030329
\citep{2003Natur.423..847H}, 031203 \citep{2006ApJ...636..967W}, 060218
\citep{2006astro.ph..3279C} -- shows that the three limits for SGRBs
substantially overlap with low-redshift LGRBs.

Models of short GRBs from neutron star (NS) mergers
\citep{2003MNRAS.345.1077R} seem naturally to produce wide opening angles
for the neutrino-annihiliation driven outflow unless the baryonic wind from
the remnant exerts significant confinement
\citep{2003MNRAS.343L..36R,2005A&A...436..273A}.
However, such wide opening angles could be problematic for the total
energy released in such a scenario unless the efficiency is fairly high.
Magnetic mechanisms may therefore be a more likely candidate to provide the
energy release in NS-NS mergers
\citep{2006Sci...312..719P,2005ApJ...630L.165L}.

It is interesting to note that GRB\,000301C, suggested to be a SGRB
\citep[duration 2\,s with a hard spectrum,][]{2001A&A...370..909J}, has an
opening angle at the extreme end of the distribution for LGRBs
\citep[$12\pm1\degr$,][]{2006ApJ...637..889Z}, as well as a strong (1
magnitude) deviation from a power-law decay at 4--5 days after the trigger.
But at the same time, GRB\,000301C is at a fairly high redshift, $z=2.04$,
much further away than the known SGRB redshifts. At a much lower redshift,
the duration of this burst would lie well within the SGRB range.
GRB\,000301C also has a damped Ly$\alpha$ (DLA) absorption system and
extinction detected in its afterglow spectrum \citep{2001A&A...370..909J},
suggesting an actively star-forming galaxy. Two other bursts are worth
noting in this discussion: GRB\,001025A, an IPN-localised hard burst with
duration 2.9\,s \citep{2006ApJ...636..381P}, and GRB\,060206, also a hard
burst with duration 7\,s \citep{2006GCN..4697....1P}. In the case of
GRB\,001025A it had a fast decay and no detected optical afterglow to a
limit of $R>25.5$ at $1.2$ days
\citep{2002A&A...393L...1W,2006ApJ...636..381P}. For GRB\,060206, the
redshift is high \citep[$z=4.05$][]{2006A&A...451L..47F} -- at low redshift,
this GRB would have had a duration about as short as GRB\,050724. It also
shows huge variability in the optical
\citep{2006astro.ph..2495S,2006astro.ph..3181M,2006ApJ...642L..99W}. Like
GRB\,000301C, its spectrum also has a DLA absorption system
\citep{2006A&A...451L..47F}.

The strong variations observed in almost all short GRBs where there is even
a moderate coverage of the lightcurve, make it difficult to determine breaks
in the power-law decays. Indeed, there is a possibility that some
lightcurves may be dominated by flaring, with little of the flux contributed
by an underlying power-law decay. In cases with long-duration, large
amplitude flaring, jet-break times would not be determined, leading to very
large opening angle limits. However this explanation of large opening angles
seems unlikely in most cases and is contradicted by the detection of a
clean, relatively slow power-law decay in the optical in GRB\,050709 and in
the X-ray in GRB\,051221A.

\section{Conclusions}
The SGRB 050709 was the first GRB with a detected optical afterglow
\citep{2005Natur.437..859H}. It is the only SGRB where clear evidence for a
jet break in the optical lightcurve has been claimed. We have shown that the
optical decay of this GRB follows a single steep power-law decay with a
rebrightening at $\sim 10$ days. There is no evidence of a jet break. The
optical rebrightening in GRB\,050709 is not easily compatible with models
involving supernovae, shock heating of a stellar companion or
non-uniformities in the accretion disk. So far there is no compelling
evidence for a jet break in any SGRB and available limits are not compatible
with the distribution of opening angles in long GRBs. There is no strong
evidence for collimation in short bursts, implying that short GRBs may be
more energetic than previously believed.

\begin{acknowledgements}
We are grateful to Andreas Zeh for providing the opening angle
determinations for long GRBs. The Dark Cosmology Centre is funded by the
DNRF. We acknowledge benefits from collaboration within the EU FP5 Research
Training Network, `Gamma-Ray Bursts: An Enigma and a Tool'.
\end{acknowledgements}

\end{document}